\documentclass{aa}  

\usepackage{graphicx}
\usepackage{txfonts}
\usepackage[utf8]{inputenc}
\usepackage[T1]{fontenc}
\usepackage[english]{babel}
\usepackage{graphicx}
\graphicspath{{./figures/}}
\usepackage{color}
\usepackage{booktabs}
\usepackage{tabularx}
\usepackage{multicol}
\usepackage{multirow}
\usepackage{hyperref}

\makeatletter
\renewcommand*\aa@pageof{, page \thepage{} of \pageref*{LastPage}}
\makeatother
\hypersetup{%
    colorlinks=false,linktoc=all, pdfstartpage=1, pdfstartview=FitV, pdfencoding=auto,%
    breaklinks=true, pdfpagemode=UseNone, pageanchor=false,
    pdfpagemode=UseOutlines,%
    plainpages=false, bookmarksnumbered, bookmarksopen=true, bookmarksopenlevel=2,%
    hypertexnames=true, pdfhighlight=/O,
    urlcolor=webbrown, linkcolor=RoyalBlue, citecolor=webgreen,
}
\def\aap{A\&A}

\def\apjs{ApJS}
\def\apjl{ApJL}

\newcommand{\squeze}{\texttt{SQUEZE}}
\newcommand{\rf}{\texttt{RF}}
\newcommand{\lgbm}{\texttt{LGBM}}
\newcommand{\cnn}{\texttt{CNN}}
\newcommand{\cnnne}{\texttt{CNN1NE}}
\newcommand{\ann}{\texttt{ANN}}
\newcommand{\qpz}{\texttt{QPz}}
\newcommand{\lephare}{\texttt{LePhare}} 

\begin{document}

   \title{The miniJPAS survey quasar selection}

   \subtitle{V. combined algorithm}

   \date{Received XXXX; accepted YYYY}
   
   \author{
    Ignasi Pérez-Ràfols\inst{1,2}\fnmsep\thanks{E-mail: \href{mailto:ignasi.perez.rafols@upc.edu}{ignasi.perez.rafols@upc.edu}}
    % Raul, Mat, Natália, Ginés i companyia
    \and L. Raul Abramo\inst{3}
    \and Gin\'es Mart\'inez-Solaeche\inst{4}
    \and Nat\'alia V.N. Rodrigues\inst{3}
    \and Matthew M. Pieri\inst{5}
    \and Marina Burjalès-del-Amo\inst{1}
    \and Maria Escolà-Gallinat\inst{1}
    \and Montserrat Ferré-Abad\inst{1}
    \and Mireia Isern-Vizoso\inst{1}
    \and Jailson Alcaniz\inst{6}
    \and Narciso Benitez\inst{7}
    \and Silvia Bonoli\inst{8,9}
    \and Saulo Carneiro\inst{6}
    \and Javier Cenarro\inst{9}
    \and David Cristóbal-Hornillos\inst{9}
    \and Renato Dupke\inst{6}
    \and Alessandro Ederoclite\inst{9}
    \and Rosa María González Delgado\inst{4}
    \and Siddhartha Gurung-Lopez\inst{10, 11}
    \and Antonio Hernán-Caballero\inst{9}
    \and Carlos Hernández–Monteagudo\inst{12,13}
    \and Carlos López-Sanjuan\inst{9}
    \and Antonio Marín-Franch\inst{9}
    \and Valerio Marra\inst{10, 11, 12}
    \and Claudia Mendes de Oliveira\inst{17}
    \and Mariano Moles\inst{9}
    \and Laerte Sodré Jr.\inst{17}
    \and Keith Taylor\inst{18}
    \and Jesús Varela\inst{9}
    \and Héctor Vázquez Ramió\inst{9}
}
\institute{
    Departament de Física, EEBE, Universitat Politècnica de Catalunya, c/Eduard Maristany 10, 08930 Barcelona, Spain
    \goodbreak
    \and
    Institut de Física d’Altes Energies (IFAE), The Barcelona Institute of Science and Technology, Campus UAB, 08193 Bellaterra Barcelona, Spain
    \and
    Departamento de F\'isica Matem\'atica, Instituto de F\'{\i}sica, Universidade de S\~ao Paulo, Rua do Mat\~ao 1371, CEP 05508-090, S\~ao Paulo, Brazil
    \and
    Instituto de Astrof\'isica de Andaluc\'ia (CSIC), P.O. Box 3004, 18080 Granada, Spain
    \and
    Aix Marseille Univ, CNRS, CNES, LAM, Marseille, France
    \and
    Observatório Nacional, Rua General José Cristino, 77, São Cristóvão, 20921-400, Rio de Janeiro, RJ, Brazil
    \and
    Independent Researcher
    \and
    Donostia International Physics Center (DIPC), Manuel Lardizabal Ibilbidea, 4, San Sebastián, Spain
    \and
    Centro de Estudios de Física del Cosmos de Aragón (CEFCA), Plaza San Juan, 1, E-44001, Teruel, Spain
    \and
    Departamento de Física, Universidade Federal do Espírito Santo, 29075-910, Vitória, ES, Brazil
    \and
    INAF -- Osservatorio Astronomico di Trieste, via Tiepolo 11, 34131 Trieste, Italy
    \and
    IFPU -- Institute for Fundamental Physics of the Universe, via Beirut 2, 34151, Trieste, Italy  
    \and
    Observatori Astron\`omic de la Universitat de Val\`encia, Ed. Instituts d’Investigaci\'o, Parc Cient\'ific. C/ Catedr\'atico Jos\'e Beltr\'an, n2, 46980 Paterna, Valencia, Spain
    \and
    Departament d’Astronomia i Astrof\'isica, Universitat de Val\`encia, 46100 Burjassot, Spain
    \and
    Instituto de Astrofísica de Canarias, C/ Vía Láctea, s/n, E-38205, La Laguna, Tenerife, Spain
    \and
    Universidad de La Laguna, Avda Francisco Sánchez, E-38206, San Cristóbal de La Laguna, Tenerife, Spain
    \and
    Departamento de Astronomia, Instituto de Astronomia, Geofísica e Ciências Atmosféricas, Universidade de São Paulo, São Paulo, Brazil
    \and
    Instruments4, 4121 Pembury Place, La Canada Flintridge, CA 91011, U.S.A.
    }

% \abstract{}{}{}{}{} 
% 5 {} token are mandatory
 
  \abstract
   {}
   {Quasar catalogues from narrow-band photometric data are used in a variety of applications, including targeting for spectroscopic follow-up, measurements of supermassive black hole masses, or Baryon Acoustic Oscillations. Here, we present the final quasar catalogue, including redshift estimates, from the miniJPAS Data Release constructed using several flavours of machine-learning algorithms.}
   {In this work, we use a machine learning algorithm to classify quasars, optimally combining the output of 8 individual algorithms.
   We assess the relative importance of the different classifiers. We include results from 3 different redshift estimators to also provide improved photometric redshifts. We compare our final catalogue against both simulated data and real spectroscopic data. Our main comparison metric is the $f_1$ score, which balances the catalogue purity and completeness.
   }
   {We evaluate the performance of the combined algorithm using synthetic data. In this scenario, the combined algorithm outperforms the rest of the codes, reaching $f_1=0.88$ and $f_1=0.79$ for high- and low-z quasars (with $z\geq2.1$ and $z<2.1$, respectively) down to magnitude $r=23.5$. We further evaluate its performance against real spectroscopic data, finding different performances. We conclude that our simulated data is not realistic enough and that a new version of the mocks would improve the performance. Our redshift estimates on mocks suggest a typical uncertainty of $\sigma_{\rm NMAD} =0.11$, which, according to our results with real data, could be significantly smaller (as low as $\sigma_{\rm NMAD}=0.02$). We note that the data sample is still not large enough for a full statistical consideration.}
   {}


   \maketitle

\section{Introduction}
In astronomy, there is now an ever-increasing number of surveys producing increasingly larger amounts of data. For instance, using multi-slit/multi-fibre spectrographs mounted on large telescopes can now obtain thousands of spectra in a single night. Examples of present and future quasar spectroscopic surveys include the 5th generation of the Sloan Digital Sky Survey (SDSS-V; \citealt{Kollmeier+2017, Almeida+2023}), the Dark Energy Spectroscopic Instrument (DESI; \citealt{Levi+2013, DESI2016a, DESI2016b, DESI2024}), the Subaru Prime Focus Spectrograph (PFS, \citealt{Takada+2014}), the Multi-Object Optical and Near-infrared Spectrograph (MOONSM \citealt{Cirasuolo+2011}), the 4-metre Multi-Object Spectroscopic Telescope (4MOST; \citealt{deJong+2019}), Gaia \citep{Storey-Fisher+2024}, the William Herschel Telescope Enhanced Area
Velocity Explorer Collaboration (WEAVE; \citealt{Dalton+2016}).

Targets for these spectroscopic surveys are drawn from even larger photometric surveys. These surveys observe many more objects, but the derived properties, such as object classification and redshifts, are less secure. Examples of present and future photometric surveys include the DESI Legacy Imaging Survey \citep{Dey+2019}, the Square Kilometer Array (SKA; \citealt{Dewdney+2009}), the Dark Energy Survey (DES; \citealt{DES2016}), the Panoramic Survey Telescope and Rapid Response System(Pan-STARRS; \citealt{Chambers+2016}), Euclid \citep{Euclid2024}, the Large Synoptic Survey Telescope (LSST; \citealt{Ivezic+2019}), and the Javalambre Physics of the Accelerating Universe Astrophysical Survey (J-PAS; \citealt{Benitez+2014}).

Amongst other interesting objects, these surveys compile lists of quasars, distant, very powerful galaxies powered by accretion into supermassive black holes at the centre of their host galaxies. Their (optical) spectra present several broad emission lines and an approximate power-law continuous emission in the UV and optical wavelengths. The quasar spectra as a whole are strikingly self-similar (e.g. \citealt{Merloni+2016} and references therein) and have been widely studied (e.g. \citealt{Jensen+2016, Lyke+2020, Wu+2022}, Pérez-Ràfols et al. 2025).

While quasars themselves are plenty interesting, they are also often used as background sources to study the intergalactic medium, and through it, cosmology. For example, the Lyman $\alpha$ absorption seen in the spectra of quasars has been widely used to constrain cosmology via the Baryon Acoustic Oscillations (see e.g. \citealt{DESI2024IV} and references therein). Other uses of the Lyman $\alpha$ absorption include intensity mapping studies (e.g. \citealt{Ravoux+2020, Kraljic+2022}, and references therein). The spectra of quasars also allow us to detect a variety of absorption systems, including Damped Lyman $\alpha$ absorbers, containing most of the neutral hydrogen in the universe (e.g. \citealt{Brodzeller+2025} and references therein) and, in general, galaxies in absorption (e.g. \citealt{Morrison+2024, Perez-Rafols2023b}).

From the many applications, here we are interested in the quasar identification and redshift estimation in photometric surveys. Both machine learning and template fitting have been used in this endeavour (see e.g. \citealt{Salvato+2018} for a review). When machine learning is used, there exists a large number of approaches to distinguish between galaxies, stars and quasars (see e.g. \citealt{Krakowski+2016, Bai+2019, Logan+2020, Xiao-Qing+2021, He+2021}). One of the main differences between the different algorithms is the input parameters that are fed to the machine learning algorithms: fluxes, magnitudes, ... More recently, even colours measured in different apertures are used \citep{Saxena+2024}.

Similarly, previous papers in this series introduced different flavours of machine-learning algorithms to construct a quasar catalogue from the miniJPAS survey \citep{Rodrigues+2023, Martinez-Solaeche+2023, Perez-Rafols+2023}. The miniJPAS survey is the first Data Release from the JPAS Collaboration and shows the results of their pathfinder camera \citep{Bonoli+2020}. In these papers, we trained our models, and assessed their performance, using dedicated mocks \citep{Queiroz+2022}. In this work, we present the final combined quasar catalogue, including the results of all previous papers in the series. As inputs for the model, we do not use the observed object properties, but rather the results of previous algorithms. 

The novelty of this work resides not only in the combination of different algorithms to build classification confidence but also in a much better assessment of the results. Besides assessing the performance with mocks, we additionally test our algorithms against real data from the DESI Early Data Release \citep{DESI2024}.

This paper is organised as follows. In section~\ref{sec:data} we describe the data used in this work, including the individual algorithms from the previous papers in the series. We then describe the combination procedure in section~\ref{sec:combine}, and assess its performance against mocks in section~\ref{sec:performance}. In section~\ref{sec:quasar_cat} we present our quasar catalogue, and we discuss its performance against real data in section~\ref{sec:discussion}. We finish summarizing our conclusions in section \ref{sec:summary}. Throughout this work, reported magnitudes refer to the AB magnitude system.

\section{Data}\label{sec:data}

\subsection{miniJPAS data \& mocks}\label{sec:minijpas_data}
This work is centred on the classification of sources from the miniJPAS Data Release \citep{Bonoli+2020}. This is a photometric survey using 56 filters, of which 54 are narrow band filters with FWHM $\sim 14$\AA{} and 2 are broader filters extending to the UV and the near-infrared. The SDSS broadband filters $u$, $g$, $r$ and $i$ complement the aforementioned 56 filters to reach a total of 60 filters. miniJPAS was run on the AEGIS field and covers $\sim1{\rm deg}^{2}$.

Sources are identified using the \texttt{SEXTRACTOR} code \citep{Bertin+1996}. There are several catalogues available according to the filter(s) used to identify the sources. Here we use the dual mode, in which objects are identified in both the filter of interest and the reference filter ($r$-band). We refer the reader to \cite{Bertin+1996, Bonoli+2020} for more detailed explanations of the software and object detection.
We clean the initial catalogue\footnote{Available on \url{https://archive.cefca.es/catalogues/minijpas-pdr201912}}, with 64,293 objects, to exclude flagged objects. These flags indicate problems during field reduction and/or object detection and are described in detail in \cite{Bonoli+2020}. Our final clean sample contains 46,441 objects. We label this sample as "all objects".

At this point, it is worth noting that high-redshift quasars are typically point-like sources. Thus, we define our main sample, which we label as "point-like", to include only point-like objects. These are found using the stellarity index constructed from image morphology using Extremely Randomised Trees (ERT, \citealt{Baqui+2021}). Following \cite{Queiroz+2022, Rodrigues+2023, Martinez-Solaeche+2023, Perez-Rafols+2023}, we require objects to be classified as stars (point-like sources) with a probability of at least 0.1, defined in their catalogue as $ERT\geq0.1$. In cases where the ERT classification failed (identified as ERT=99.0), we instead use the alternative classification using the stellar-galaxy locus classifier from \cite{Lopez-Sanjuan+2019}, requiring a minimum probability of $SGLC\geq0.1$. 11,419 objects meet this point-like source criterion and constitute our point-like sample.

In addition to miniJPAS data, we also use simulated data (mocks). These mocks were already used in the previous papers on the series to train different algorithms (see section~\ref{sec:individual_algorithms}). Here, we will use them as truth tables to (optimally) combine said algorithms. The mocks are based on SDSS spectra, which are convolved with the J-PAS filters. Noise is also added to match the expected signal-to-noise. We refer the reader to \cite{Queiroz+2022} for a detailed description of the mocks and their building procedure, but we note that the mocks were defined to match the properties of the point-like objects in miniJPAS. We have a total of 360,000 objects distributed between the train (100,000), validation (30,000) and test (30,000) sets. These samples contain an equal number of stars, quasars and galaxies. This is not the expected on-sky distribution of sources but it is convenient in order to train the algorithms. Thus, in addition to these samples, we have a special 1 deg$^{2}$ test set built in a similar procedure except that it has the expected relative fraction for each type of object, containing 2,191 stars, 6,410 galaxies and 510 quasars. Again, we refer the reader to \cite{Queiroz+2022} for details on how this sample is built.

In the previous papers in the series, where we trained the different classifiers we used here (see section~\ref{sec:individual_algorithms}), we restricted the analysis to r-band magnitudes $17.0<r\leq24.3$. Thus, we need to apply the same cut here, which reduces the number of objects in all our samples. Table~\ref{tab:samples} summarises the final number of objects in each sample, when including this magnitude cut. 

\begin{table*}
    \centering
    \caption{Summary of the samples used in this work. The first column specifies the type of sample (mocks or real data). The second and third columns give the name of the sample and the number of objects it contains. Where available, in the following columns, we also provide the number of stars, galaxies, and quasars separately. Note that the numbers provided here include a magnitude cut of $17.0<r\leq24.3$ (see text for details).}
    \label{tab:samples}
    \begin{tabular}{llcccc}
        \toprule
        type & sample & objects & star & galaxy & quasar \\
        \midrule
        \multirow{4}{*}{real data} 
        & all objects & 40,805 & $\dots$ & $\dots$ & $\dots$ \\
        & point-like & 6,654 & $\dots$ & $\dots$ & $\dots$\\
        & DESI cross-match & 3730 & 667 & 2873 & 190\\
        & DESI cross-match point-like & 996 & 480 & 357 & 159\\
        \midrule
        \multirow{4}{*}{mocks} 
        & training & 297,959 & 99,931 & 99,109 & 98,919\\
        & validation & 29,783 & 9,991 & 9,901 & 9,891 \\
        & test & 29,779 & 9,995 & 9,897 & 9,887 \\
        & test 1 deg$^{2}$ & 9,036 & 2,187 & 6,347 & 502 \\
        \bottomrule
    \end{tabular}
    .
\end{table*}

\subsection{DESI data}\label{sec:desi_data}
One of the major drawbacks of our previous papers is that we relied solely on mocks to assess the performance of our classifiers. True, there were SDSS spectra with miniJPAS counterparts, but they were too few (272) to extract any statistically meaningful information. Fortunately, the DESI Early Data Release (EDR, \citealt{DESI2024}) is now available to use. We find 3730 objects with miniJPAS counterparts. From these, the DESI automatic classification pipeline \texttt{Redrock} (Bailey et al. in prep) finds 667 stars, 2900 galaxies and 163 quasars, of which 53 have redshift $z\geq2.1$ and 110 have $z<2.1$. However, according to the DESI Visual Inspection (VI) program, performed during the Survey Validation phase, \texttt{Redrock} sometimes fails in its classification \citep{Alexander+2023, Lan+2023}. Because we are using this classification as a truth table, and because of the relatively small sample size, we visually inspected all the spectra and fixed a small number of classifications (see Appendix~\ref{sec:vi}). We found 491 stars, 2873 galaxies and 190 quasars, of which 133 with $z\geq2.1$ and 57 with $z<2.1$. We label this sample as "DESI cross-match".

As we did with our main catalogue, we create a smaller "DESI cross-match point-like" catalogue containing only the point-like objects, defined according to miniJPAS data and using the criteria mentioned in section~\ref{sec:minijpas_data}. There are a total of 1,171 objects split into 655 stars, 357 galaxies and 159 quasars, of which 111 have $z\geq2.1$ and 48 have $z<2.1$. We note that there is a non-negligible fraction ($\sim18\%$) of quasars that are classified as extended sources by miniJPAS. From these, 9 are high-z quasars and 22 are low-z quasars. At small redshifts, we often see the galactic emission, and therefore we expect some of them to be classified as extended sources. However, this should not be the case for high-z objects. These mislabelling as extended sources are at the faint end of our magnitude distribution, and this is likely the reason behind the misclassification.

Finally, a few of the objects lie outside our selected magnitude range (see section~\ref{sec:minijpas_data}) and are thus discarded from this analysis. This magnitude cut is applied based on the miniJPAS magnitudes associated with these objects (irrespective of the measured DESI magnitude). The final number of objects in each of the samples is summarised in the first bloc of table~\ref{tab:samples}.

\subsection{Previous classifiers}\label{sec:individual_algorithms}
Contrary to the other papers in the series, we do not directly use miniJPAS data for the classification, but rather the outputs from other codes. We use the outputs from 10 algorithms, of which 8 are classification algorithms (one of which also provides redshift estimates), and 2 do not perform classification but provide only redshift estimates. The classifiers are summarised in Table~\ref{tab:previous_classifiers}, and we describe them in more detail in the following paragraphs.

\begin{table*}
    \centering
    \caption{Summary of the previous classifiers included in this work. A single * in the output columns represents the different classes and should be replaced by {\it star}, {\it gal}, {\it lqso} (for low-z quasars) and {\it hqso} (for high-z quasars). In the same column, the double ** represents alternative classifications and should be replaced with numbers 0 to 4 (0 being the most confident classification and 4 being the least confident). We refer the reader to the respective papers for details about the architectures and hyperparameters of the codes.}
    \label{tab:previous_classifiers}
    \scalebox{0.85}{
    \begin{tabular}{lp{0.13\textwidth}lp{0.2\textwidth}p{0.2\textwidth}p{0.2\textwidth}}
        \toprule
        Name & Algorithm used & Type & Input & Output & Authors \\
        \midrule
        CNN1 & Convolutional Neural Network & Classification only & Array of fluxes and errors & \texttt{conf\_*\_CNN1} & \cite{Rodrigues+2023}\\
        CNN1NE & Convolutional Neural Network & Classification only & Array of fluxes & \texttt{conf\_*\_CNN1NE} & \cite{Rodrigues+2023} \\ 
        CNN2 & Convolutional Neural Network & Classification only & 2D image including fluxes and errors & \texttt{conf\_*\_CNN2} & \cite{Rodrigues+2023} \\
        RF & Random Forest & Classification only & Array of fluxes & \texttt{conf\_*\_RF} & \cite{Rodrigues+2023} \\
        LGBM & LightGBM & Classification only & Array of fluxes & \texttt{conf\_*\_LGBM} & \cite{Rodrigues+2023} \\
        ANN1 & Artificial Neural Network & Classification only & Array of magnitudes + hybridisation & \texttt{conf\_*\_ANN1} & \cite{Martinez-Solaeche+2023} \\
        ANN2 & Artificial Neural Network & Classification only & Array of fluxes + hybridisation & \texttt{conf\_*\_ANN2} & \cite{Martinez-Solaeche+2023} \\
        SQUEzE & Random Forest & Classification and redshift & Specific line metrics & \texttt{conf\_SQUEZE\_**}, \texttt{z\_SQUEZE\_**}, \texttt{class\_SQUEZE} & \cite{Perez-Rafols+2023} \\
        QPz & Template-based fitting & Redshift only & Array of fluxes & \texttt{zphot\_QPz} & Queiroz, private communication \\
        LePhare & Template-based fitting & Redshift only & Array of fluxes & \texttt{zphot\_LePhare} & Queiroz, private communication \\
        \bottomrule
    \end{tabular}
    }

\end{table*}

The first 5 algorithms, called \cnn1{}, \cnnne{}, \cnn2{} \rf{}, and \lgbm{}, are defined in \cite{Rodrigues+2023}. The first three are based on Convolutional Neural Networks (CNNs). \cnn1{} and \cnnne{} use the fluxes as 1D arrays of input parameters to perform the classification. The difference between them is that \cnn1{} also includes the flux errors, whereas \cnnne{} does not. \cnn2{} also uses the fluxes and their errors but instead of formatting them as 1D arrays, a 2D matrix representation is built. The other two algorithms, \rf{} and \lgbm{}, use different flavours of Decision Tree to perform the classification, the Random Forest and the Light Gradient Boosting Machine, respectively. Details on the implementation and performance of these algorithms are given in \cite{Rodrigues+2023}. They focus on classification using 4 categories: star, galaxy, low-z quasar ($z<2.1$), and high-z quasar ($z\geq2.1$), and do not provide redshift estimates. We note that the $z=2.1$ pivot was chosen because the Lyman $\alpha$ feature enters the optical range at this redshift. Each of the algorithms outputs 4 quantities for each analysed object corresponding to the confidence of classification for each class. For example, \texttt{conf\_lqso\_CNN1} gives the confidence of objects being low redshift quasars according to the classification by \cnn1{}. They also output the object's class by selecting the class with maximal confidence. We do not include these columns in our training set as they only contain redundant information.

The following 2 algorithms are described in \cite{Martinez-Solaeche+2023}. They use Artificial Neural Networks to classify the object based on the J-PAS magnitudes (\ann{}1) or the J-PAS fluxes (\ann{}2). Hybridisation in the training samples is used to prevent the classifiers from becoming overly confident. The output of these codes is the same as the previous 5, and they do not provide redshift estimates.

The last code that performs classification is \squeze{}. \squeze{} was originally developed to run on spectra \citep{Perez-Rafols+2020b,Perez-Rafols+2020a} and was later adapted to running on miniJPAS data \citep{Perez-Rafols+2023}. Its behaviour differs significantly from the previous codes. First, the code looks for emission peaks in the spectra and assigns a set of possible emission lines to each of them. This generates a list of trial redshifts, $z_{\rm try}$. For each of these trial redshifts, a set of metrics are computed around the predicted position of the quasar emission lines, which are then fed to a random forest algorithm that determines which of the trial redshifts corresponds to the true quasar redshifts. For each of the possible classifications (the trial redshifts), \squeze{} provides the confidence of the object being a quasars and a redshift estimate. We refer the reader to \cite{Perez-Rafols+2020b} for a more detailed description of \squeze{}. Normally, for each object, the best classification (the one with the largest confidence) is kept, and the rest are discarded. However, there is information present in the subsequent classifications (for example, if more than one emission line is correctly identified, the subsequent classifications will also have high confidence and a similar redshift estimate). Thus, here we keep the best 5 classifications, which we label as \squeze{}\_0 to \squeze{}\_4. Each of the classifications contains two variables: the classification confidence, e.g. \texttt{conf\_SQUEZE\_0}, and the classification redshift, e.g. \texttt{z\_SQUEZE\_0}. In addition, we include the final classification of \squeze{} in terms of the usual 4 categories: star, galaxy, low-z quasar and high-z quasar, based on the confidence and redshift of \squeze{}\_0. We label this variable \texttt{class\_SQUEZE}.

Finally, to complement the redshift estimations from \squeze{} we include photometric redshift estimates from \qpz{} and \lephare{} (Queiroz et al. in prep.). These are run on all objects, assuming they are quasars. The redshift estimation is performed using quasar templates, and the output variables are named \texttt{zphot\_QPz} and \texttt{zphot\_LePhare}, respectively.

\section{Combination procedure}\label{sec:combine}
As we have seen in the previous papers of the series, each of the classification algorithms can provide a quasar catalogue on its own. However, these catalogues are not the same but have slight differences. All the classifiers will agree on the ``easier'' quasars\footnote{Here, ``easier" is ill-defined. It broadly refers to quasars with a larger number of emission lines and/or higher signal-to-noise.}. Other, less clear objects have different classifications depending on the algorithm used. These differences lie in the different flavours of machine learning algorithms used by the individual classifiers. The question is, then, how to (optimally) combine the results from these algorithms to maximise the purity and completeness of the final quasar catalogue. 

Maximising either the purity or the completeness is rather easy. If we wish to maximize completeness, it suffices to compile all the objects that are in any of the individual catalogues. Similarly, to maximize the purity, we can keep all the quasars present in all the catalogues. Finding a good compromise between the two is delicate, and we use a second layer of machine learning to solve this optimization problem. 

The classification is made using a Random Forest Classifier, which combines multiple Decision Trees (a structure where the algorithm makes predictions by splitting the data set based on constraints imposed in terms of the features). Each tree is built with a subsample of the data, using the bootstrap aggregating ({\it bagging}; \citealt{Breiman+1996}) technique. Note that we explored using several other machine-learning algorithms for this classification and found that the Random Forest Classifier has the best performance. 

We implemented the model using the \texttt{scikit-learn PYTHON} package \citep{Pedregosa+2011}. We included all the features described in section~\ref{sec:individual_algorithms} except for those belonging to \qpz{} and \lephare{} (see section~\ref{sec:feature_importance}). Training is performed using the validation set, and we test our performance on the test and test 1 deg$^2$ mock samples.  This is necessary as we require the results of the individual classifiers (which are trained using the training sample) as inputs for our model.

In addition to the object classification, we train a Random Forest Regressor, which operates similarly to the Random Forest Classifier but for continuous variables, to refine the redshift estimation of the found quasars. The redshift regressor is trained using only the quasars present in the validation sample and applied only to objects classified as quasars. For objects not classified as quasars, we simply assign NaN to their redshift estimate. We implemented the model using the \texttt{xgboost PYTHON} library \citep{XGBOOST}, which allows us to use custom metrics when training the random forests (see below). We included all the features described in section~\ref{sec:individual_algorithms}. %except for those belonging to \qpz{} and \lephare{} (see section~\ref{sec:feature_importance}).

The standard implementation of the Random Forest Regressor uses the mean squared error \citep{Breiman+2001}. Because the errors are squared, the mean squared error metric is sensitive to outliers and penalises large deviations in the model. This works well when the distribution of errors follows (or is close to) a Gaussian distribution. In our case, the error distribution in the photometric redshifts is inherently non-Gaussian. For instance, for bright objects, the error is dominated by line confusion, making a small set of outcomes more likely to be selected. What is more, beyond a certain threshold in redshift error, we do not care about the error itself, and the estimate is labelled as catastrophic (e.g. \citealt{Bolzonella+2020}).

Indeed, we found that using the standard metric was unsuitable for our regression problem. Instead, we used the normalised median absolute deviation (see section~\ref{sec:redshift_precision}) as our training metric. In addition, we noted that for bright objects (with $r<22$), the random forest estimates were not as good as the individual estimation by \qpz{}. Therefore, we decided to keep the \qpz{} estimates for these objects and apply our estimator only to fainter objects.

\section{Performance of the classifiers}\label{sec:performance}
We now assess the performance of the combined algorithm and compare it to the performance of the individual algorithms. We start by assessing the classification performance in section~\ref{sec:classification_performance} and move to analyse the importance of the different classifiers in section~\ref{sec:feature_importance}. We then discuss the redshift precision in section~\ref{sec:redshift_precision}. The results are summarised in table~\ref{tab:results}.

\begin{table*}
    \centering
    \caption{Summary of the performance for the different classifiers. Metrics are given including objects brighter than $r=23.5$. For each classifier, the first row corresponds to metrics for the high-z quasars (with $z \ge 2.1$) and the second row corresponds to metrics for the low-z quasars (with $z < 2.1$). The first bloc corresponds to the performance on mocks (on the test and test 1 deg$^2$ samples), and the second bloc corresponds to the performance on the DESI cross-match data (excluding and including the extended sources).}
    \label{tab:results}
    \scalebox{0.85}{
\begin{tabular}{p{0.1\textwidth}|cccccc|ccccccccc}
\toprule
\multirow{2}{*}{Name} & \multicolumn{3}{c}{Test} & \multicolumn{3}{c|}{Test 1 deg$^2$} & \multicolumn{3}{c}{DESI cross-match (point-like)} & \multicolumn{3}{c}{DESI cross-match (all)} \\
& $f_1$ score & purity & comp.& $f_1$ score & purity & comp.& $f_1$ score & purity & comp.& $f_1$ score & purity & comp.\\
\midrule
\squeze{} & 0.60 & 0.70 & 0.52 & 0.37 & 0.28 & 0.54 & 0.69 & 0.65 & 0.73 & 0.42 & 0.30 & 0.66 & \\
& 0.37 & 0.51 & 0.29 & 0.17 & 0.12 & 0.29 & 0.60 & 0.68 & 0.53 & 0.29 & 0.21 & 0.47 & \\
\cnn1{} & 0.87 & 0.91 & 0.83 & 0.77 & 0.71 & 0.84 & 0.65 & 0.55 & 0.82 & 0.27 & 0.16 & 0.82 & \\
& 0.76 & 0.77 & 0.76 & 0.41 & 0.28 & 0.75 & 0.64 & 0.58 & 0.73 & 0.41 & 0.30 & 0.64 & \\
\cnnne{} & 0.75 & 0.84 & 0.68 & 0.65 & 0.58 & 0.73 & 0.73 & 0.73 & 0.73 & 0.51 & 0.39 & 0.74 & \\
& 0.67 & 0.65 & 0.69 & 0.30 & 0.19 & 0.67 & 0.63 & 0.53 & 0.78 & 0.26 & 0.16 & 0.69 & \\
\cnn2{} & 0.81 & 0.91 & 0.73 & 0.77 & 0.76 & 0.78 & 0.67 & 0.63 & 0.73 & 0.30 & 0.19 & 0.72 & \\
& 0.73 & 0.73 & 0.73 & 0.36 & 0.24 & 0.73 & 0.70 & 0.63 & 0.77 & 0.45 & 0.34 & 0.67 & \\
\rf{} & 0.64 & 0.58 & 0.71 & 0.33 & 0.22 & 0.72 & 0.60 & 0.60 & 0.59 & 0.28 & 0.18 & 0.62 & \\
& 0.59 & 0.54 & 0.65 & 0.20 & 0.12 & 0.64 & 0.55 & 0.43 & 0.75 & 0.25 & 0.16 & 0.68 & \\
None & 0.74 & 0.82 & 0.67 & 0.61 & 0.52 & 0.74 & 0.72 & 0.74 & 0.70 & 0.50 & 0.39 & 0.70 & \\
& 0.66 & 0.64 & 0.67 & 0.28 & 0.18 & 0.63 & 0.63 & 0.55 & 0.75 & 0.31 & 0.20 & 0.67 & \\
\ann1{} & 0.77 & 0.86 & 0.70 & 0.73 & 0.67 & 0.81 & 0.78 & 0.80 & 0.75 & 0.60 & 0.50 & 0.74 & \\
& 0.71 & 0.70 & 0.73 & 0.36 & 0.24 & 0.74 & 0.61 & 0.50 & 0.79 & 0.26 & 0.16 & 0.72 & \\
\ann2{} & 0.74 & 0.86 & 0.65 & 0.63 & 0.56 & 0.72 & 0.78 & 0.80 & 0.75 & 0.58 & 0.47 & 0.76 & \\
& 0.70 & 0.66 & 0.74 & 0.30 & 0.19 & 0.71 & 0.60 & 0.48 & 0.78 & 0.25 & 0.16 & 0.71 & \\
Combined & 0.88 & 0.91 & 0.85 & 0.81 & 0.76 & 0.88 & 0.62 & 0.52 & 0.77 & 0.24 & 0.14 & 0.78 & \\
algorithm & 0.79 & 0.79 & 0.79 & 0.44 & 0.30 & 0.82 & 0.66 & 0.60 & 0.73 & 0.43 & 0.33 & 0.64 & \\
\bottomrule
\end{tabular}
}
\end{table*}

\subsection{Classification performance}\label{sec:classification_performance}
We start by assessing the classification performance. Our main scoring metric for the comparison is the $f_1$ score metric (but we also provide the purity and completeness). This metric is convenient as it combines purity and completeness in a single metric:
\begin{equation}
    f_{1} = \frac{2pc}{p+c} ~.
\end{equation}
This metric helps us balance having high values of purity and completeness. When any of the two becomes too low, then the overall $f_1$ score drops. We compare the $f_1$ score of the different classifiers as a function of the limiting magnitude, i.e. including all the objects brighter than a given magnitude. This will give us an idea of the faintest magnitude to which we can trust the resulting catalogue. 

\begin{figure*}
    \centering
    \includegraphics[width=\textwidth]{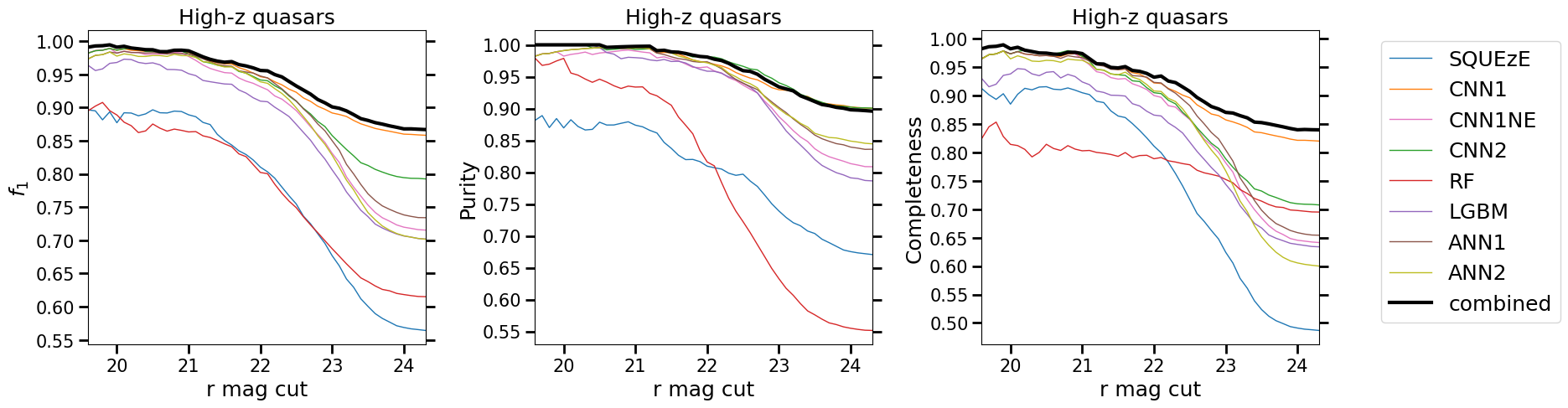}\\
    \includegraphics[width=\textwidth]{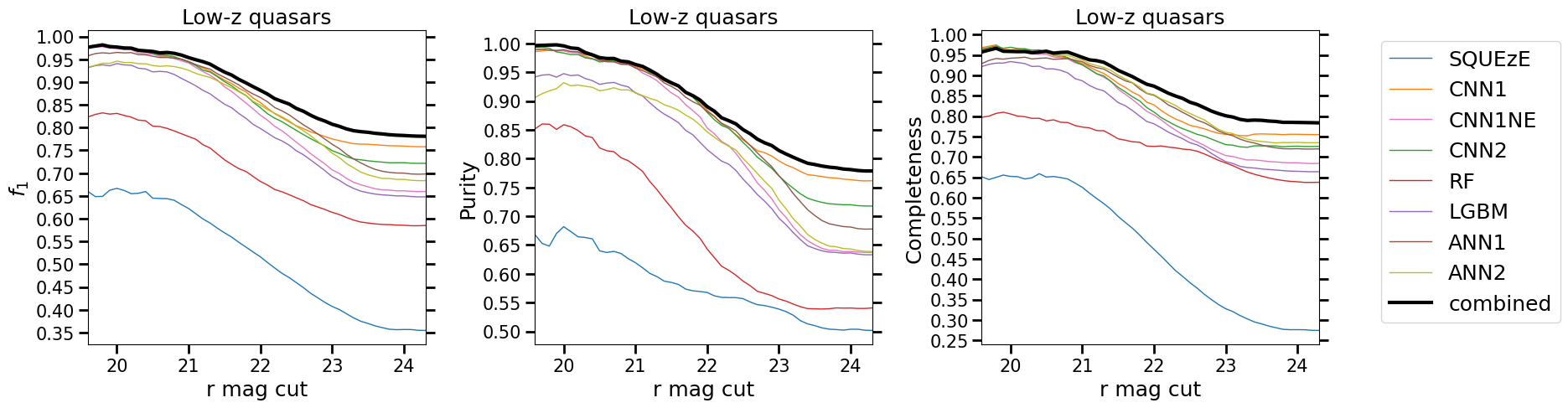}
    \caption{$f_1$ score  (left), purity (middle) and completeness (right) versus limiting magnitude for the test sample. The thick solid lines show the performance of the combined algorithm, to be compared with the performance of the individual classifiers (thin lines). The top panel shows the performance for high-z quasars (with $z\geq2.1$) and the bottom panel the performance for low-z quasars (with $z<2.1$). We see that the combined algorithm outperforms the rest.}
    \label{fig:f1_vs_mag}
\end{figure*}

Figure~\ref{fig:f1_vs_mag} shows the $f_1$ score, purity and completeness as a function of limiting magnitude for the quasar classes for the test sample. The two classes are high-z quasars (with $z\geq2.1$) and low-z quasars (with $z<2.1$)\footnote{Note that we expect the classification of high-z quasars to be easier as they typically contain a larger number of emission line features.}. The results from the combined algorithms are shown as a thick black line, and we plot the results from the individual classifiers as thin lines of different colours. As expected, we see that the combined algorithm outperforms the rest of the algorithms.

\begin{figure*}
    \centering
    \includegraphics[width=\textwidth]{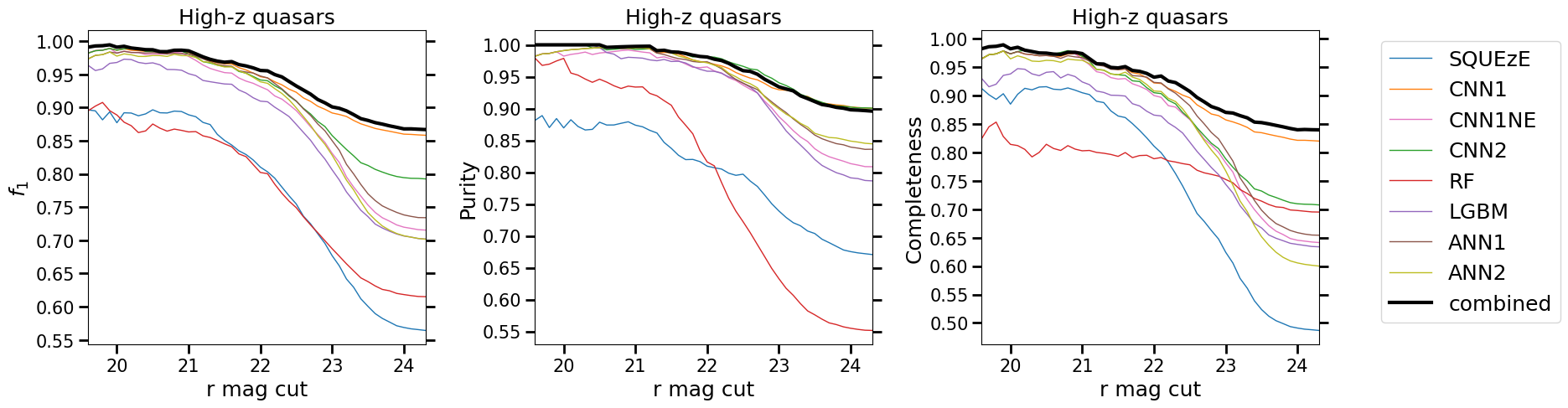}
    \includegraphics[width=\textwidth]{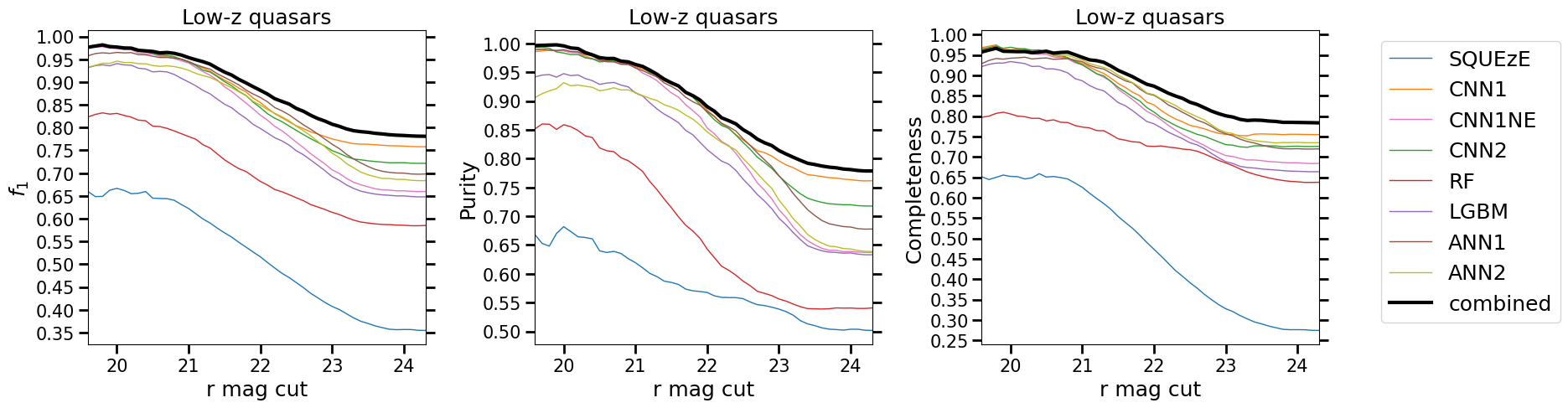}
    \caption{Same as figure~\ref{fig:f1_vs_mag} but for the test 1 deg$^2$ sample}
    \label{fig:f1_vs_mag_test1deg2}
\end{figure*}

We compute the $f_1$ score, purity and completeness down to magnitudes 24.3 because that is the faintest magnitude in our samples. However, we note that we have very few objects at these very faint magnitudes\footnote{While the number of objects increases with magnitude, this is only true down to the limiting magnitude of the survey (included in our mocks). Only a small number of objects will go beyond this limit due to noise.}, which explains the flatness of the curves at the faint end. For high-z-quasars, and considering magnitudes brighter than $r=23.5$ (where we still have a decent number of objects per unit magnitude), we find an $f_1$ score of 0.88 (corresponding to a purity of 0.91 and a completeness of 0.85) for the combined algorithm, and we need to decrease the magnitude limit to $r=23.2$ or brighter to achieve $f_1$ scores greater than 0.9 (corresponding to a purity of 0.93 and a completeness of 0.86). Similarly, for low-z quasars, we find $f_1=0.79$ (corresponding to a purity and a completeness of 0.79) at a magnitude cut of $r=23.5$, and we need to decrease to a magnitude cut of $r=21.8$ to achieve a score of 0.9 (corresponding to a purity of 0.91 and a completeness of 0.89).

One of the main issues with the test sample is that it contains a balanced number of stars, galaxies and quasars. In real data, we do not expect to find this distribution. Instead, we expect there to be fewer quasars. To test the impact of this, we analyse the performance of the algorithm on the test 1 deg$^2$ sample, with relative numbers of stars, galaxies, and quasars matching the expectations on real data. We show the results of this exercise in figure~\ref{fig:f1_vs_mag_test1deg2}, where we see a similar trend but with slightly lower $f_1$ scores driven by a decrease in the purity. However, the completeness levels are roughly constant. For high-z quasars, we find $f_1=0.81$ (corresponding to a purity of 0.76 and a completeness of 0.88) for a magnitude limit of $r=23.5$, and we need to decrease the magnitude limit to $r=22.7$ to achieve scores greater than 0.90 (corresponding to a purity of 0.89 and a completeness of 0.90). For low-z quasars, we find $f_1=0.44$ (corresponding to a purity of 0.30 and a completeness of 0.82) at a magnitude cut of $r=23.5$, and we need to decrease to a magnitude cut of $r=20.0$ to achieve a score of 0.93 (corresponding to a purity of 0.91 and a completeness of 0.94). From this, we conclude that our high redshift sample is more secure and less affected by the number of galaxies. This is expected as low-redshift quasars have a diffuse physical difference with galaxies (there are cases where we can see both the central AGN emission and the galactic emission).

\subsection{Feature importance}\label{sec:feature_importance}
To better understand the classifications made by the combined algorithm, we use a modified version of the permutation feature importance technique to estimate the importance of the different features. Briefly, when applying the standard permutation feature importance we shuffle the values of every feature, one at a time, and observe the resulting degradation of the model's performance \citep{Breiman+2001}. This shuffling breaks the relationship between feature and target and allows us to determine how much the model relies on this particular feature. We shuffle each feature $j$ a total $N=10$ times, obtaining the scores $s_{k,j}$ ($k=1..N$). The final importance is computed by subtracting the average scores of the different runs from the reference score $s$:
\begin{equation}
    i_j = s - \frac{1}{N}\sum_{k=1}^{N}s_{k,j} ~. 
\end{equation}

While this technique is very model-agnostic, it may result in misleading values when features are strongly correlated, as is the case here. For instance, for a given classifier the confidences add up to 1. Thus, instead of shuffling a single feature, we shuffle blocks of features, grouped by the classifier. In our final classifier, we include the information about the classification algorithms. Instead, when studying feature importance here, we also include the redshift estimations from \qpz{} and \lephare{}. We also study the addition of the r-band magnitude as a feature, which was found to be important for the individual algorithms.

Feature importance can be estimated using multiple scorers. In our case, we use the $f_1$ score as our main indicator, but we also provide the importance with respect to the purity and completeness. We do this separately for both high-z and low-z quasars.

\begin{figure*}
    \centering
    \includegraphics[width=\textwidth]{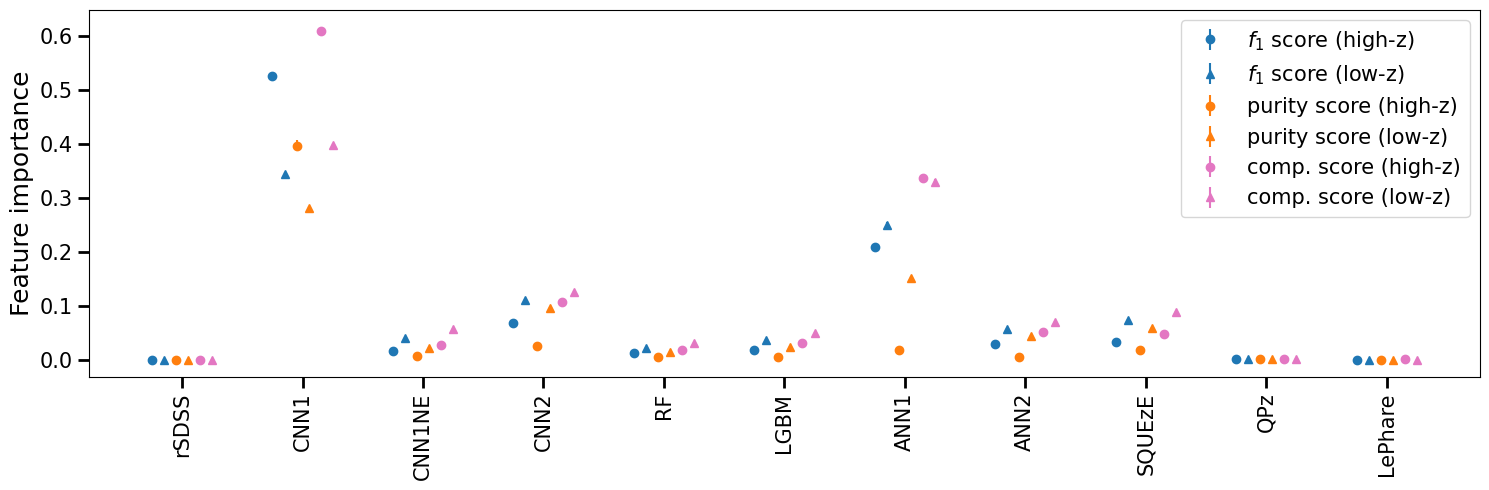}
    \caption{Feature importance analysis of the combined algorithm classifier based on a modified version of the permutation feature importance technique (see text for details). The y-axis shows the feature importance for different scorers ($f_1$ score, purity and completeness), and for high-z and low-z quasars separately. These should be compared with the performance values listed in table~\ref{tab:results}. Columns $rSDSS$, \qpz{} and \lephare{} are not included in our final classifier.}
    \label{fig:feature_importance_class}
\end{figure*}

Results from this exercise, considering objects brighter than $r=23.5$, are given in figure~\ref{fig:feature_importance_class}. We can see that the three most important classifiers are, in order of importance, the \cnn1{}, the \ann1{} and the \cnn2{}. We also see that the r-band magnitude and the redshift estimation from \qpz{} and \lephare{} are not relevant and are therefore excluded from our final classifier. 

\subsection{Redshift precision}\label{sec:redshift_precision}
We now analyse the precision of our redshift estimate. We remind the reader that, as described in section~\ref{sec:combine}, in normal operations, we will only compute redshift estimates for objects classified as quasars. However, here we will analyse the results of applying the redshift estimator to all the quasars of the test sample, irrespective of their classification. This will help disentangle the effect of mistakes in the classification from the actual precision of our redshift estimates.

\begin{figure}
    \centering
    \includegraphics[width=0.48\textwidth]{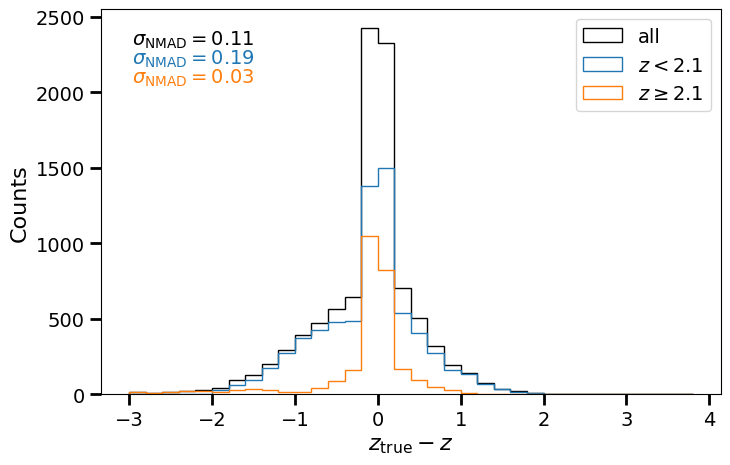}
    \caption{Distribution of redshift errors for the test sample. The distribution, including all objects, is given in the black line. We also include the distributions for the high/low redshift split, respectively. The split is performed at $z=2.1$ and considering the predicted redshift values.}
    \label{fig:z_distribution}
\end{figure}

Figure~\ref{fig:z_distribution} shows the distribution of the error in redshift given as $\Delta z = z_{\rm true} - z$, where $z_{\rm true}$ is the `true' redshift of the object, given in the truth table, and $z$ is our redshift estimate. Besides the overall distribution, in figure~\ref{fig:z_distribution}, we also show the distributions for high-z and low-z quasars separately, where the split is done based on the predicted redshifts (as opposed to using the preferred class). 

All three distributions are centred around zero, and we do not observe any systematic shift. However, we note a non-negligible amount of catastrophic errors in z, particularly for the low-z quasars. The redshift estimation for high-z quasars is typically more secure thanks to the presence of multiple emission lines, reducing the line confusion problem (i.e. correctly identifying an emission line but assigning the wrong label and, therefore, a catastrophically different redshift)

To estimate the typical redshift uncertainty from this distribution, we use the normalised median absolute deviation, $\sigma_{\rm NMAD}$, defined by \cite{Hoaglin+1983} as
\begin{equation}
    \sigma_{\rm NMAD} = 1.48\times {\rm median}{\left(\frac{\left|z_{\rm true}-z\right|}{1+z_{\rm true}}\right)}~.
\end{equation}
We obtain a value of $\sigma_{\rm NMAD}=0.11$ for the entire sample, $\sigma_{\rm NMAD}=0.03$ for high-z quasars and $\sigma_{\rm NMAD}=0.19$ for low-z quasars. Indeed, this confirms that the redshift precision is much higher for high-z quasars. Again, having multiple emission lines helps secure the redshift information. 

We compare the performance of the combined algorithm with that of the individual classifications: \squeze{}, \qpz{} and \lephare{}. For the comparison with \squeze{}, we take its best classifications, i.e. \texttt{z\_SQUEZE\_0}, and we obtain $\sigma_{\rm NMAD}=0.25$, $0.27$, and $0.19$ for the entire sample, low-z quasars, and high-z quasars, respectively. Similarly, for \qpz{} we obtain $\sigma_{\rm NMAD}=0.21$, $0.21$, and $0.20$, respectively, and for \lephare{} we obtain $\sigma_{\rm NMAD}=0.83$, $0.84$, and $0.006$, respectively. We see that, generally, the combined algorithm outperforms the rest of the algorithms. The exception to this is the obtained value of $\sigma_{\rm NMAD}$ for high-z quasars using \lephare{}. We note, however, that this comes at the cost of \lephare{} having a significantly worse performance for low-z quasars, which translates to an overall worse performance.

\begin{figure}
    \centering
    \includegraphics[width=0.48\textwidth]{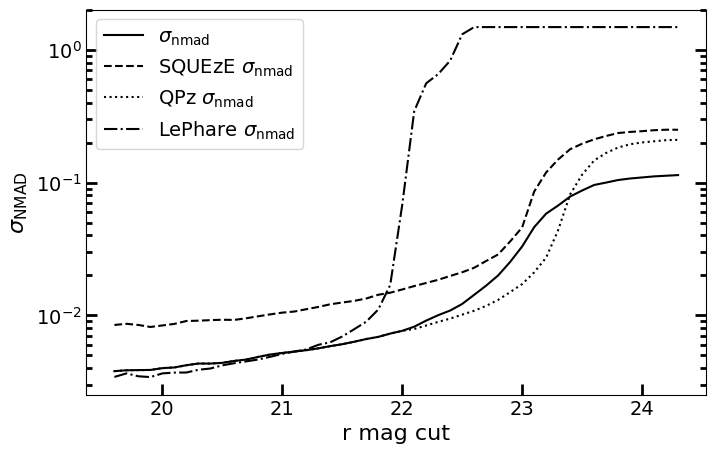}\\
    \includegraphics[width=0.48\textwidth]{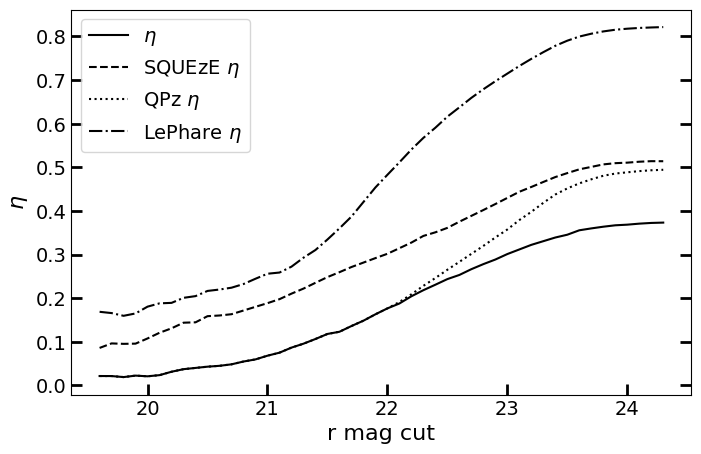}\\
    \caption{Top panel: Evolution of the measured $\sigma_{\rm NMAD}$ as a function of limiting magnitude. The solid line shows the results for our combined algorithm. The other lines show the measured $\sigma_{\rm NMAD}$ for \qpz{} (dotted), the best estimate from \squeze{} (dashed), i.e. \texttt{Z\_SQUEZE\_0}, and \lephare{} (dash-dotted). The bottom panel shows the fraction of outliers, i.e. objects with $\left|\Delta z\right|/(1 + z_{\rm true})>0.15$, also as a function of limiting magnitude. We can see that, while the recovered $\sigma_{\rm NMAD}$ for the combined algorithm is sometimes larger than that of \qpz{} this is compensated by a decrease in the outlier fraction.}
    \label{fig:z_std_vs_mag}
\end{figure}

We further explore the evolution of the redshift error with the limiting magnitude in figure~\ref{fig:z_std_vs_mag}. In the top panel, we see that for magnitudes fainter than $r=22$, the measured $\sigma_{\rm NMAD}$ (solid line) is generally larger than the corresponding line obtained when analysing \qpz{} results (dotted line), except when we include the faintest objects (with $r\gtrsim23.5$). This increase in $\sigma_{\rm NMAD}$ can be explained by the inclusion of the other, less precise, redshift estimates. The benefit of adding these estimates can be seen in the bottom panel of figure~\ref{fig:z_std_vs_mag}, showing a consistent decrease in the outlier fraction, i.e. the fraction of catastrophic redshift classifications (defined as objects with $\left|\Delta z\right|/(1 + z_{\rm true})>0.15$). In other words, the alternative redshift estimates, while not as precise as \qpz{}, help alleviate the line confusion problem. As we mentioned before, for objects brighter than $r=22$ we choose to keep \qpz{} estimates as we otherwise see a similar increase in $\sigma_{\rm NMAD}$ without the benefit of decreasing the outlier fraction

We finalize this section by performing a feature importance analysis on our redshift classifier. We follow the procedure described in section~\ref{sec:feature_importance} except that we change the scoring functions. Because the scoring function needs to be maximised (i.e. larger scores are better), we choose to use $-\sigma_{\rm NMAD}$ as our scorer. We take three scorers: one including all objects, one including high-z quasars only and one including low-z quasars.

\begin{figure*}
    \centering
    \includegraphics[width=\textwidth]{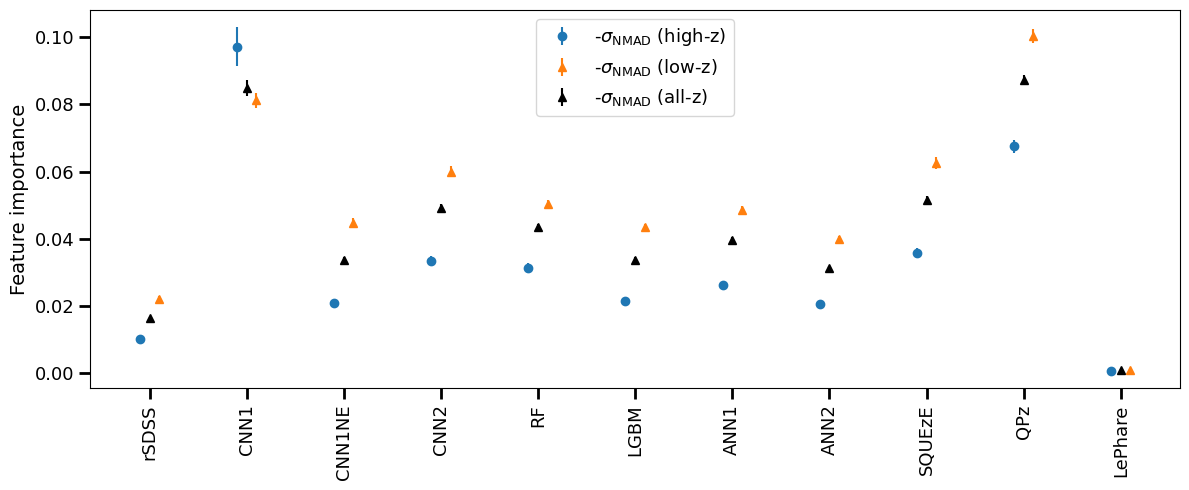}
    \caption{Feature importance analysis of the combined algorithm redshift estimator based on a modified version of the permutation feature importance technique (see text for details). The y-axis shows the feature importance for the $-\sigma_{\rm NMAD}$ scorer. Different colours show the score for the entire sample (black), high-z quasars (orange) and low-z quasars (blue). These should be compared to the measured $\sigma_{\rm NMAD}=0.11$, $0.03$ and $0.19$, respectively.}
    \label{fig:feature_importance_z}
\end{figure*}

Results are given in figure~\ref{fig:feature_importance_z}. We see that the dominant features are the redshift estimate from \qpz{} and the classifications by \cnn1{}, followed by the estimates from \squeze{}. We also see that \lephare{} estimates are the least relevant.

\section{Quasar catalogues}\label{sec:quasar_cat}
In Section~\ref{sec:performance}, we have seen the performance of the combined algorithm on mock data. Now we apply it to the Mini J-PAS sources to build our final catalogues. For some applications, a more complete or purer catalogue will be needed. Therefore, in our catalogues, we include all objects classified as quasars (irrespective of the classification confidence).

We built two catalogues. The first one includes only the point-like sources, including 784 sources. The second one also includes the extended sources and includes 11,487 sources. Here, it is worth stressing that the performance discussed in section~\ref{sec:performance} applies to the point-like sources. The unreasonably large number of objects when including extended sources indicates that we cannot extend the usage of the combined algorithm beyond its training set, at least not in a straightforward manner. We provide the catalogue, including extended sources, as a bonus and at the user's risk. %However, we will discuss tentative performances for both catalogues using real data in section~\ref{sec:discussion}. 

\begin{figure*}
    \centering
    \includegraphics[width=0.48\textwidth]{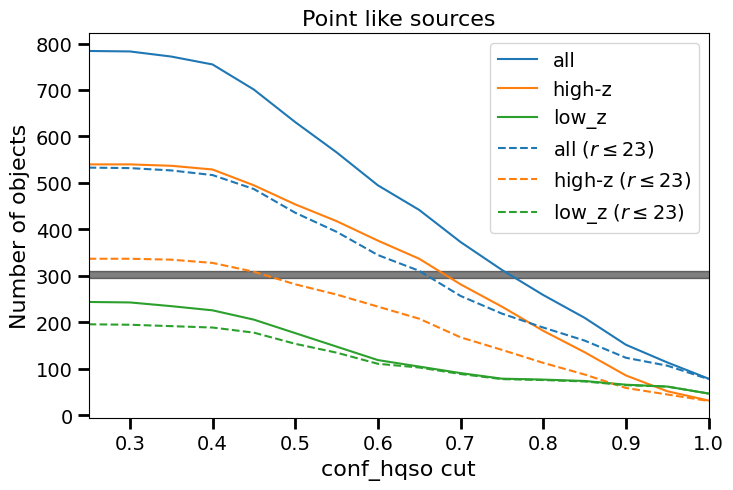}
    \includegraphics[width=0.48\textwidth]{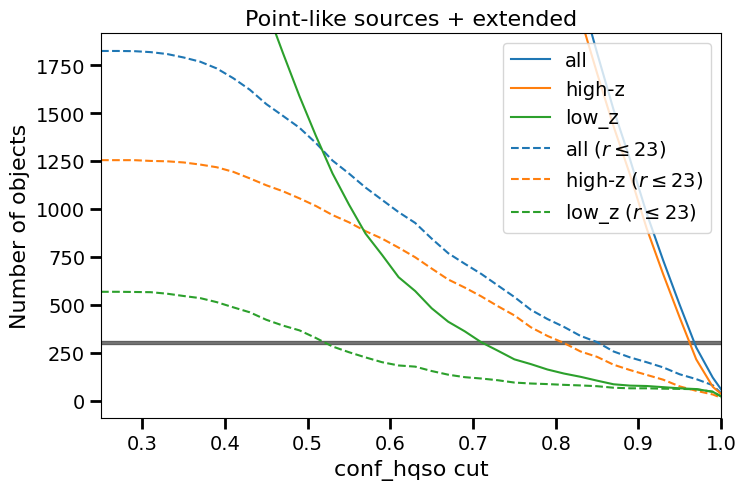}
    \caption{Number of entries in our quasar catalogues as a function of classification confidence. The orange and green lines include the high-z and low-z entries, respectively. The blue lines are the sum of the orange and green lines. Solid lines include all the entries, irrespective of their magnitude, whilst the dashed lines exclude objects fainter than $r=23$.
    Grey horizontal bands indicate the expected number of quasars down to magnitude $r=23$ according to the models from \protect\cite{Palanque-Delabrouille+2016}. The left panel shows the number of objects for the point-like catalogue and the right panel also includes the extended sources. We note that in the bottom panel, the number of objects including all magnitudes is unreasonably large, and thus we choose to focus the y-label on the magnitude-limited lines.}
    \label{fig:catalogue_numbers}
\end{figure*}

We show the number of quasars as a function of classification confidence in figure~\ref{fig:catalogue_numbers}. We take as classification confidence the highest confidence between the high-z and low-z quasar classes. As expected, the number of quasars decreases with the classification confidence\footnote{Naturally, a minimum confidence of 0.25 arises given that we are choosing from 4 classes}. 
Analysing this figure, one can immediately see that there are way too many entries, especially when extended sources are included. This may suggest problems in our mocks that would impact the expected performance (this is further discussed in section~\ref{sec:discussion}). We can, however, limit the number of objects by imposing confidence thresholds. 

Results from \cite{Palanque-Delabrouille+2016} predict the number of quasars up to magnitude $r=23$ to be between 296 and 311, depending on the chosen luminosity model, of which between 81 and 91 are expected to be high-z quasars (see their table 5). Note that the numbers shown here include objects that go significantly fainter than this. If we restrict to objects brighter than $r=23$, we are left with 520 entries in the point-like catalogue and 1,769 when including the extended sources. For the point-like sample, imposing a confidence cut $\geq0.67$ reduces these numbers to a more sensible 298 quasars, of which 81 are high-z quasars. To achieve similar results when also including extended objects we need to impose a confidence cut $\geq0.87$. In this case, we would obtain 304 quasars, of which 94 have high-z.

\section{Discussion}\label{sec:discussion}
So far, we have shown that our combined algorithm outperforms the individual algorithms it was trained on. However, one of the main drawbacks of our analysis so far is that we completely rely on mocks for both training the algorithm and evaluating its performance. What is more, the number of quasars in our catalogues seems unrealistically large (albeit this issue can be mitigated by imposing a confidence cut). In previous papers of this series, we also argued that spectroscopic follow-up of the candidates is required to fully validate our findings \citep{Rodrigues+2023, Martinez-Solaeche+2023, Perez-Rafols+2023}.

\begin{figure*}
    \centering
    \includegraphics[width=\textwidth]{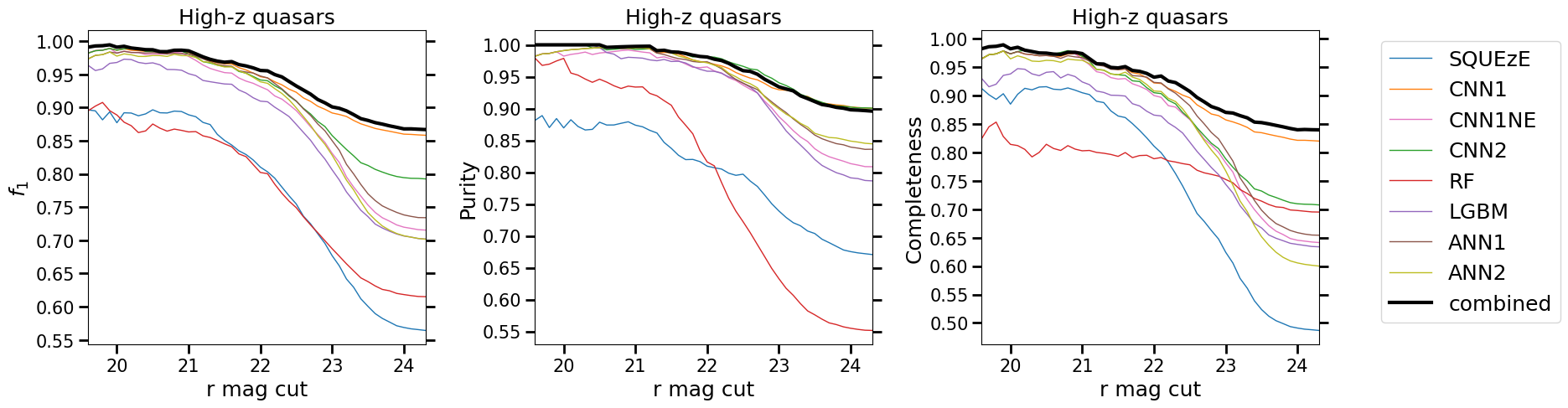}
    \includegraphics[width=\textwidth]{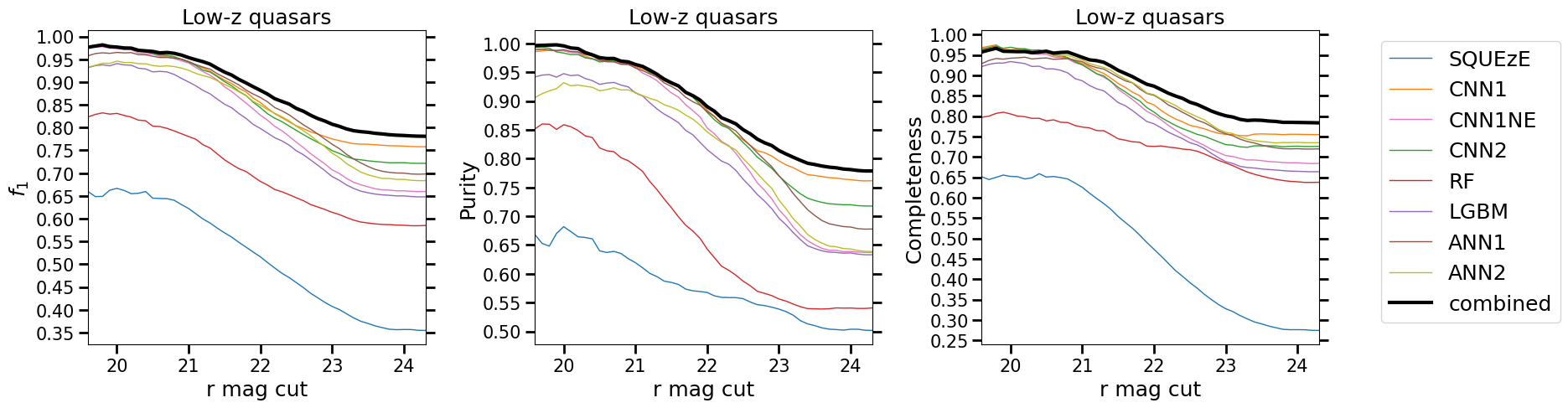}
    \caption{Same as figure~\ref{fig:f1_vs_mag} but for the DESI cross-match point-like sample.}
    \label{fig:f1_vs_mag_desi_point_like}
\end{figure*}

Fortunately, with the publication of the DESI EDR, a fraction of our candidates have spectroscopic observations (see section~\ref{sec:desi_data}) which we visually inspected to assign labels (see appendix~\ref{sec:vi}). Here, we use this sample to further probe the performance of the algorithm. We treat this sample as a truth table and compute the $f_1$ score for the classification. Results of this exercise, considering only the point-like sources, are given in figure~\ref{fig:f1_vs_mag_desi_point_like} and summarised in table~\ref{tab:results}. 

Comparing this plot with figures~\ref{fig:f1_vs_mag} and \ref{fig:f1_vs_mag_test1deg2} we immediately see clear differences in the behaviour of the combined algorithm and the other classifiers. The most evident feature is that now the combined algorithm does not outperform the other algorithms. This can be explained by a drop in the performance of the \cnn1{} algorithm, particularly concerning the achieved purity. As we saw in section~\ref{sec:performance}, the combined algorithm is dominated by the classifications by \cnn1{} (see figure~\ref{fig:feature_importance_class}). 

In addition, we see that the change in performance is very algorithm-dependent. At a magnitude cut of $r=23.5$, and compared to the performance on the test 1deg$^2$ sample, we see a decrease in the $f_1$ score for high-z quasars of 0.12 for \cnn1{}, 0.10 for the \cnn2{}, and 0.16 for the combined algorithm. For other algorithms, we see instead an increase in the performance. From larger to smaller increase we have 0.32 for \squeze{}, 0.26 for \rf{}, 0.15 for \ann2{}, 0.11 for \lgbm{}, 0.08 for \cnnne{} and 0.04 for \ann1{}. For low-z quasars, we see a general increase in the performance. Again from larger to smaller increase in the $f_1$ score, we have 0.43 for \squeze{}, 0.35 for \rf{} and \lgbm{}, 0.33 for \cnn2{} and \cnnne{}, 0.30 for \ann2{}, 0.25 for \ann1{}, 0.23 for \cnn1{} and 0.22 for the combined algorithm.

It is worth noting that our mocks were created using SDSS spectra \citep{Queiroz+2022} and that DESI data is deeper. Mock objects that are fainter than the SDSS limit are thus drawn from brighter objects to which extra noise is added. Alas, we know the quasar properties to vary with magnitude (e.g. the Baldwin effect \citealt{Baldwin+1977}). Considering all this together, these results indicate that our mocks are not realistic enough, and a refined version is necessary. The different performances of the different algorithms are also suggestive of their potential overfitting and their resilience. Algorithms like \squeze{}, which rely on higher-level metrics for the classification, are suppressed in the combined algorithm due to their worse performance in mocks but have the greatest increase in performance when confronted with real data. Thus, once a new, more refined version of the mocks becomes available and the combined algorithm is retrained, we expect to see a significantly different feature importance analysis. 

We now turn our attention to the performance of our redshift estimates. As we mentioned in section~\ref{sec:combine}, we only run our redshift estimates in objects classified as quasars. Because we have already determined that the combined algorithm should not be used in extended sources, we study the redshift precision only for the sample in the DESI cross-match point-like.

\begin{figure}
    \centering
    \includegraphics[width=0.48\textwidth]{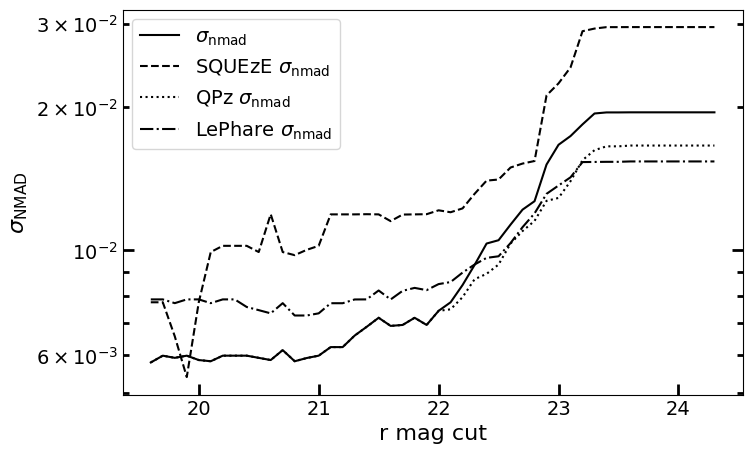}\\
    \includegraphics[width=0.48\textwidth]{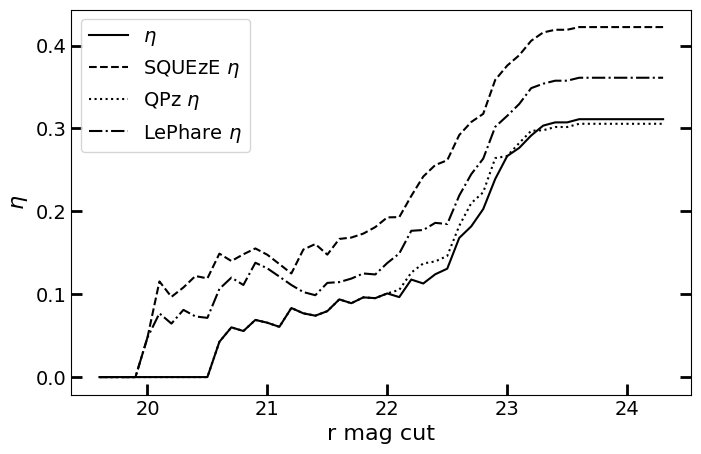}
    \caption{Same as figure~\ref{fig:z_std_vs_mag} but for the DESI cross-match point-like sample.}
    \label{fig:z_std_vs_mag_desi}
\end{figure}

We see the results of this exercise in figure~\ref{fig:z_std_vs_mag_desi}. We get an overall $\sigma_{\rm NMAD}=0.02$ to be compared with the previous value of 0.11 estimated using mocks. The fraction of outliers is also generally smaller in real data than in mocks. We again see clear differences in the behaviour of the combined algorithm and the other classifiers. One big difference when compared to the performance changes seen in the classifier is that the redshift precision is better in real data than in mocks. Nevertheless, this change in performance again suggests that our mocks are not realistic enough (as discussed above). 

\section{Summary and conclusions}\label{sec:summary}
In previous papers of the series, we have presented several machine-learning algorithms to construct quasar catalogues based on the automatic identification of the objects. Here, we have presented an extra layer of machine learning to optimally combine the results from the individual algorithms. We have then tested the algorithm on mocks and on data from the DESI EDR. We now summarise our conclusions:
\begin{itemize}
    \item When applied to mocks, the combined algorithm outperforms the rest of the classifiers. This is quantified using the $f_1$ score. For the test sample, we reach a $f_1=0.88$ ($f_1=0.79$) for high-z (low-z) quasars brighter than $r=23.5$. We need to increase the magnitude cut to $r=23.2$ and $r=21.8$, respectively, to reach $f_1=0.9$. 
    \item A feature importance analysis reveals that the combined algorithm is dominated by the \cnn1{} code, to where small adjustments are made based on the results from other codes (in particular the \cnn2{} and the \ann1{}).
    \item When applied to data observed in DESI EDR, we find significant differences in the performance. This statement holds for both the combined algorithm and the individual classifiers. We note that for some algorithms the performance increases, and for others, it decreases. In this case, the combined algorithm does not outperform the rest. We conclude that this signals an issue with the mocks themselves (and not the training of the combined algorithm). New, improved mocks are required in order to fix this issue.
    \item  We provide redshift estimates for our quasars. Based on mocks, we find a typical error of $\sigma_{\rm NMAD}=0.18$, but it seems to be much smaller than this when applied to data observed in DESI EDR, $\sigma_{\rm NMAD}=0.02$. 
    \item We provide a list of quasar candidates in the miniJPAS field with the associated confidence of classification. A confidence cut of 0.67 recovers a reasonable number of candidates.
    \item The combined algorithm was trained using point-like sources. It cannot be applied to extended data with similar performance levels. We note, however, that a few quasars are observed as extended sources. To recover them, this should be taken into account in the mocks. 
\end{itemize}

\begin{acknowledgements}

This paper has gone through an internal review by the J-PAS collaboration. Based on observations made with the JST/T250 telescope and JPCam at the Observatorio Astrofísico de Javalambre (OAJ), in Teruel, owned, managed, and operated by the Centro de Estudios de Física del Cosmos de Aragón (CEFCA). We acknowledge the OAJ Data Processing and Archiving Unit (UPAD) for reducing and calibrating the OAJ data used in this work. Funding for the J-PAS Project has been provided by the Governments of Spain and Aragón through the Fondo de Inversión de Teruel, European FEDER funding and the Spanish Ministry of Science, Innovation and Universities, and by the Brazilian agencies FINEP, FAPESP, FAPERJ and by the National Observatory of Brazil. Additional funding was also provided by the Tartu Observatory and by the J-PAS Chinese Astronomical Consortium.

IPR was supported by funding from the European Union's Horizon 2020 research and innovation programme under the Marie Sklodowskja-Curie grant agreement No. 754510. RA is supported by the FAPESP grant 2022/03426-8. GMR and RMGD acknowledge financial support from the Severo Ochoa grant CEX2021-001131-S, funded by MICIU/AEI/10.13039/501100011033. They are also grateful for financial support from project PID2022-141755NB-I00, and project ref. AST22\_00001\_Subp 26 and 11, with funding from the European Union – NextGenerationEU. MMP was supported by the French National Research Agency (ANR) under contract ANR-22-CE31-0026 and by Programme National Cosmology et Galaxies (PNCG) of CNRS/INSU with INP and IN2P3, co-funded by CEA and CNES. VM thanks CNPq (Brazil) and FAPES (Brazil) for partial financial support.

\end{acknowledgements}
    
    \bibliographystyle{aa}
    \bibliography{main}

\begin{appendix}
\section{Visual inspection of DESI EDR objects}\label{sec:vi}
We visually inspected the 3,720 DESI EDR objects with miniJPAS counterparts to search for potential flaws in the automatic classification. Each spectrum is evaluated using the \texttt{Prospect} tool\footnote{\url{https://github.com/desihub/prospect}}, which is the same tool used by DESI \citep{Alexander+2023, Lan+2023}. At least two people examined each spectrum and determined the quality of the automatic classification. Whenever there was disagreement between the inspectors, a final independent inspector examined the spectrum and made the final decision. We note that we did not find a case where the final inspector was not in agreement with one of the previous inspections. 

Since we are interested in the bulk classification between stars, galaxies, low-z quasars and high-z quasars, we did not attempt to refine the redshift estimates and only corrected them whenever we found a catastrophic error in their estimation. Of particular note is the classification of Active Galactic Nuclei (AGNs), where we also see the galactic emission. We classified them as quasars as long as we could observe the presence of a broad \ion{Mg}{II} line, irrespective of the strength of the galactic component. 

We find a total of 667 stars, 2,873 galaxies, 57 high-z quasars and 133 low-z quasars. We recover 4 high-z quasars and 23 low-z quasars. Most of the recovered low-z quasars are indeed these AGNs, where we detect the galactic emission. We note that DESI corrects the automatic classification by Redrock with the usage of afterburners to recover these objects \citep{Alexander+2023}. Having analysed this sample, we confirm that the DESI automatic pipeline performs excellently, missing only a small number of quasars.

\section{Data model for catalogue files}\label{sec:files}
Data in the quasar catalogue is provided in compressed fits files (.fits.gz) in a single Header Data Unit. For each entry, we provide both our final classification and the classification from the individual algorithms. Details on the columns are given in table~\ref{tab:files}.

\begin{table*}
    \centering
    \caption{Column description in the data format. Here, XXX should be replaced by \cnn1{}, \cnnne{}, \cnn2{}, \rf{}, \lgbm{}, \ann1{}, and \ann2{}. YYY should likewise be replaced by numbers from 0 to 4 (both included). For CLASS* variables 0=star, 1=galaxy, 2=low-z quasar (z<2.1), and 3=high-z quasar (z>=2.1). For CONF\_SQUEZE\_* and Z\_SQUEZE\_* variables the higher number indicates less confident classifications.}\label{tab:files}
    \begin{tabular}{ll}
        \toprule
        Column & Description \\
        \midrule        
        TILE\_ID & ID of the Tile image of the object \\
        NUMBER & Number ID assigned by Sextractor to object \\
        CLASS\_TRUE & Correct classification (if available) \\
        RA & Right Ascension of the object (in degrees)\\
        DEC & Declination of the object (in degrees) \\
        RSDSS & r-band APER-3 magnitude of the object\\
        TOT\_PROB\_STAR & Probability of the object being point-like \\
        ERT\_PROB\_STAR & Probability of the object being point-like\\ 
        IS\_POINT\_LIKE & 1 for point-like objects, 0 otherwise\\
        SPECID & Unique object identifier\\
        CONF\_STAR\_XXX & Confidence of object being a star (XXX) \\
        CONF\_GAL\_XXX & Confidence of object being a galaxy (XXX) \\
        CONF\_LQSO\_XXX & Confidence of object being a z<2.1 quasar (XXX)\\
        CONF\_HQSO\_XXX & Confidence of object being a z>=2.1 quasar (XXX)\\
        CLASS\_XXX & Object class (XXX) \\
        CONF\_SQUEZE\_YYY & Confidence of object being a z<2.1 quasar (\squeze{})\\
        Z\_SQUEZE\_YYY & \squeze{} redshift estimate \\
        CLASS\_SQUEZE & Object class (\squeze{}) \\
        ZPHOT\_QPZ & \qpz{} redshift estimate\\
        ZPHOT\_LEPHARE & \lephare{} redshift estimate\\
        CONF\_STAR & Final confidence of object being a star \\
        CONF\_GAL & Final confidence of object being a galaxy \\
        CONF\_LQSO & Final confidence of object being a z<2.1 quasar\\
        CONF\_HQSO & Final confidence of object being a z>=2.1 quasar\\
        CLASS & Final object class \\
        Z & Final redshift estimate \\
        \bottomrule
    \end{tabular}
\end{table*}
    \end{appendix}

\end{document}